\documentclass[lettersize,journal]{IEEEtran}
\IEEEoverridecommandlockouts
\usepackage{graphicx} 

\usepackage{xspace}
\usepackage{algorithm}  
\usepackage{algorithmic}

\usepackage{url} 
\usepackage{mathbbol}
\newcommand{\sol}{{\em SERENE}\xspace}
\usepackage{xcolor}
\usepackage{amsmath}
\usepackage{caption}

\usepackage[caption=false,font=normalsize,labelfont=sf,textfont=sf]{subfig}

\newcommand\blfootnote[1]{%
  \begingroup
  \renewcommand\thefootnote{}\footnote{#1}%
  \addtocounter{footnote}{-1}%
  \endgroup
}

\newcounter{magicrownumbers}

\newcommand{\eg}{{\it e.g.,}\xspace}
\newcommand{\ie}{{\it i.e.,}\xspace}

\newcommand{\func}{{\it Detection}\xspace} 
\newcommand{\stabb}{{SnE}\xspace} 

\newcommand{\sne}{{SnE}}

\def\bbone{\mathbb{1}}

\newcommand{\pcol}{{\it $P_c$}\xspace}

\newcommand{\h}{{$\mathcal{H}$}\xspace}
\newcommand{\m}{{$\mathcal{M}$}\xspace}
\newcommand{\cl}{{$\mathcal{C}$}\xspace}

\begin{document}
\title{\sol: A Collu\underline{s}ion R\underline{e}silient \underline{R}eplication-based V\underline{e}rificatio\underline{n} Fram\underline{e}work}

\author{Amir Esmaeili, Abderrahmen Mtibaa 
\\University of Missouri--St Louis\\
Computer Science Department\\
\{ae3wc, amtibaa\}@umsl.edu
}

\maketitle

\begin{abstract}
\blfootnote{\color{red}A version of this paper is currently submitted and under review in the IEEE Internet of Things Journal Special Issue on Augmented Intelligence of Things for Vehicle Road Cooperation Systems.}


The rapid advancement of autonomous driving technology is accompanied by substantial challenges, particularly the reliance on remote task execution without ensuring a reliable and accurate returned results. This reliance on external compute servers, which may be malicious or rogue, represents a major security threat. While researchers have been exploring {\em verifiable computing}, and replication-based task verification as a simple, fast, and dependable method to assess the correctness of results. However, colluding malicious workers can easily defeat this method. Existing collusion detection and mitigation solutions often require the use of a trusted third party server or verified tasks which may be hard to guarantee, or solutions that assume the presence of a minority of colluding servers. We propose \sol, a collusion resilient replication-based verification framework that detects, and mitigates colluding workers. Unlike state-of-the-art solutions, \sol uses a lightweight detection algorithm that detects collusion based on a single verification task. Mitigation requires a two stage process to group the workers and identifying colluding from honest workers. We implement and compare \sol's performance to {\em Staab et. al}, resulting in an average of 50\% and 60\% accuracy improvement in detection and mitigation accuracy respectively. 

\end{abstract}


\section{Introduction}
\label{Introduction}



Autonomous driving systems, which has been transforming the automotive industry and paving the way for safer and more efficient transportation, is characterized by the volume and heterogeneity of sensory data that needs increasing complexity of algorithms, real-time processing demands, and a lot of on-board and remote processing power. In additions, tasks such as image processing for real-time self driving directions can be very critical and must be accurate. Verifiable computing~\cite{walfish2015}, which aims at verifying the accuracy of the results returned by on-board or remote servers becomes very important especially for these tasks.


Verifiable computing solutions fall into one the following three main categories: (i) attaching probabilistically checkable proofs to each offloaded task to identify incorrect results with high probability~\cite{Gennaro10}, (ii) using Trusted Execution Environments (TEEs), such as Intel Software Guard executions (SGX) to ensure the integrity of computation execution and results~\cite{jauernig2020}, and (iii) redundantly requesting the execution of tasks from multiple external servers, and applying majority voting to find the correct result~\cite{canetti2011}. While proof-based, and TEE-based solutions are limited to some specific applications, and hardware dependability, replication-based methods are generic, easy to implement, and effective~\cite{Levitin2017}. However, replication-based task verification is prune to main security attack which can be executed by two or more colluding workers submit the same incorrect results, and defeat the majority voting scheme of the replication-based mechanism (\ie collusion attack)~\cite{Silaghi09}.

Most collusion resilient replication-based solutions rely on: (1) enlarging the voting pool~\cite{Kupcu15}, (2) spot checking using a set of pre-defined trusted tasks~\cite{Zhao2005}, or trusted third party server to re-execute the task~\cite{Silaghi09}, and (3) incentivizing rational servers to betray collusion~\cite{Dong17} to decrease the chance of collusion attack. While most of these existing solutions rely on trusted third parties or may prevent and not protect against collusion, recently similarity-based clustering solutions have emerged to probe the workers and identify clusters of workers to infer {\em colluding} from {\em honest} workers~\cite{canetti2011,Silaghi09,canon2010}. These solutions, howver, fail to detect and mitigate collusion when colluding nodes represent the majority in the network. To the best of our knowledge, this problem remains unexplored. 

We propose \sol, a Collu\underline{s}ion R\underline{e}silient \underline{R}eplication-based V\underline{e}rificatio\underline{n} Fram\underline{e}work. \sol is implemented on top of any task replication-based framework to continuously monitor the list of workers and detect the presence of collusion, which triggers its mitigation process to identify and isolate colluding workers in the network. \sol's detection relies on identifying two clusters of workers consistently disagreeing with each others. While this identification guarantees the presence of colluding workers however, without assumptions of the size of colluding workers or the presence of trusted third party servers (used by state-of-the-art solutions), \sol uses a three-step mitigation algorithm to partition the group of workers and identify the colluding ones.   

The three main contribution of our paper are summarized as follows:

\begin{itemize}
    \item We propose \sol to detect and mitigate the collusion attack. \sol can accurately identify and isolate {\em colluding} nodes even when they represent 90\% of the worker population (we assume that there is a least two honest workers in the network), without relying on any trusted servers or pre-checked tasks. 
    
    \item \sol decouples detection, and mitigation phases to run the mitigation approach once the presence of colluding workers in the network is detected. While \sol's detection algorithm is periodic and consciously monitor the workers behavior, it is designed to be lightweight and does not require costly lookups. 
    
    \item We evaluate the \sol performance with state of art Staab \& Angel, which we refer to as \stabb~\cite{Staab09}, and the results show that mitigation accuracy of \sol is more than double that of \stabb. Furthermore, \sol detects collusion 15\% faster than \stabb and 30\% to 60\% more accurate detection. Finally, we perform a set of benchmarking tests to assess the run time of \sol and show that it performs faster that \stabb in three tested platform while incurring slightly more resource utilization.
\end{itemize}

The rest of this paper categorized as follows: Section~\ref{related_work} presents the literature review of verifiable computing solutions and compares state-of-the-art collusion defense methods. In the Section~\ref{system_model}, we briefly present our system model, threat model, and the assumptions we used in this paper. \sol's detection and mitigation algorithms are discussed in Section~\ref{solution}. We evaluate \sol's performance and present our simulation results in Section~\ref{Evaluation}. Challenges/discussions and concluding remarks are presented in Section~\ref{limitation} and Section~\ref{colclusion} respectively. 

\section{Related Work} \label{related_work}

Verifying the correctness of remote workers execution has been recently investigated, and the proposed research solutions can be categorized into three main categories: (i) {\em Proof-based} approaches using probabilistically checkable proofs (PCPs) produced by workers, and attached to the results~\cite{Gennaro10, Backes13, Elkhiyaoui16}, (ii) {\em TEE-based} solutions that at the hardware level guarantees codes and data integrity for task offloaders. The Intel SGX~\cite{schuster2015}, and ARM TrustZone~\cite{duarte2018} are two main TEE-based technologies, and (iii) {\em replication-based} methods that are very generic. In this approach, clients assign tasks to multiple workers, and final results are specified by a quorum (\eg 
 majority voting). A research has reduced the overhead of redundant communication by limiting task replicas~\cite{canetti2011,Chen2021}. 

Although replication-based solutions are simple, and easy to use, it is still vulnerable to colluding workers that can change majority voting. While most of researches on collusion detection mechanism are focused on replication-based methods, there are a few number of papers that have worked on collusion of TEE-based, and proof-based solutions. Proof-based collusion is limited to
the risk of delegating verification or proof setup to a colluding
third party~\cite{wang2022}, whereas the collusion for TEE-based solutions
is mainly due to a rogue remote attestation~\cite{menetrey2022}.

However, replication-based collusion defense mechanisms mainly divided into two different areas: Prevention~\cite{Belenkiy08,Kupcu15,Zhao2005,Watanable09,Dong17,Kong19}, or detection and mitigation~\cite{Silaghi09,Staab09,canon2010,Araujo11,bendahmane2014}. While in prevention solutions the main target is to incentivize colluding workers to betray the collusion~\cite{Dong17} (by givng higher rewards in a game), or enlarging the voting pool to decrease the probability of winning of colluders in majority voting~\cite{Kupcu15}, nevertheless, the main weakness of prevention is it works like a vaccine for diseases and it can not guarantee to prevent from collusions.

On the other hand, detection, and mitigation is an approach to find the colluding workers, and makes them isolated from the workers population. Silaghi et al.~\cite{Silaghi09} used two-step algorithms to identify a
majority pool of servers which will be considered benign and
used for detecting colluding servers, while Staab et al.~\cite{Staab09}
applied a one-step graph clustering algorithm to identify both
benign and colluding servers. Both of these approaches need time to complete set of works, and then applied detection algorithm, also percentage of colluding servers should be less than 50\%. Moreover, algorithms to cut the worker's graph has additional overhead. 

Zhao et al.~\cite{Zhao2005}, proposed spot checking solution to evaluate workers by sending tasks with already known results, and finding colluding servers. Spot checking tasks should be unpredictable, and it is hard to find these tasks. Some of solutions in this area assume a trust third party~\cite{sauber2021novel} for result verification, and some of them suppose there are multiple pre-arranged spot tasks~\cite{Levitin2017}. All of two assumption can be rarely used in the real world.

Unlike most of these proposed solutions, \sol does not use trusted parties (\eg workers), or trusted pre-approved tasks, and can accurately detect and mitigate collusion even when the percentage of colluding workers exceeds 50\% of the workers in the network.





\section{System Model, Threat Model, and Assumption} \label{system_model}

\subsection{System Model}
We consider an untrustworthy edge computing network consisting of $N$ workers nodes\footnote{In this paper, we use workers, servers, and edge nodes interchangeably}, $S_1 \ldots S_N$ (\eg edge computing servers) scattered in an area where clients can choose one or many workers at a time to outsource their computation tasks. We assume that clients perform replication-based verification by selecting a voting pool consisting of $k$ random workers\footnote{Without loss of generality, we use $k=3$ throughout the paper}. Users send tasks to a voting pool at any given time, and collect the majority vote to ensure task execution correctness. 
 
\subsection{Threat Model}

The system threat model includes three types of workers which submit results to a given offloaded task:

\begin{itemize}
    \item {\em honest workers:} A worker is called honest if it executes and returns correct results to any given task. We assume that honest workers can, rarely, return incorrect results with probability $\epsilon$, due to hardware failure, or incompatibility. 
    The list of honest workers is denoted by \h. 
    
    \item  {\em naive malicious workers:} This type of workers are malicious and always submit a random incorrect result (\eg do not execute the task and return any random result to same energy, or resources) for a given task. These workers work independently without any coordination with other malicious workers. The list of naive malicious workers is denoted by \m. 
    
    \item {\em colluding workers:} Colluding workers perform sophisticated attacks; they coordinate (\eg exchange command and control messages) and collude only if they count for the majority of the verification pool. In addition, they coordinate and decide to collude randomly to avoid verification detection. We denote the list of colluding workers by \cl. A colluding worker $w \in $ \cl colludes if and only if (1) it ensure there exists enough workers in the pool to form a majority, and (2) all colluding workers in the pool decide to collude with a fixed probability \pcol. We also assume that colluding workers follow an \textit{evasive} strategy, consisting of storing a list of previously seen tasks and if they a receive a task twice they assume it's a verification task and act as an {honest} worker for that task by returning the correct result. 

\end{itemize}



\subsection{Assumptions} We assume the followings: 
\begin{itemize}
    \item Clients have insufficient resources and are incapable to execute the task themselves, thus incapable of verifying the correctness of the results,
    \item Routers, switches, and clients are not malicious. The integrity of all messages exchanges is not compromised (\eg no man in the middle attack).  
    \item All colluding workers at the network edge implement the same strategy--they coordinate and agree on returning the same incorrect results with the same {\em fixed} probability of collusion $P_c$.
    
\end{itemize}



\section{\sol: Collusion Detection and Mitigation} \label{solution}

Our Collusion Resilient Replication-based Verification Framework, \sol, implements two main modules; (i) a module to detect the presence of servers' colluding behavior, and (ii) a collusion mitigation module.

\subsection{\sol's Detection Module} \label{detection-module-sub}
We design \sol's collusion detection to be lightweight and fast in detecting any potential collusion of edge servers a user is communicating with. 

Users store a set of collusion verification tasks (CVT) selected randomly from their genuine tasks--\ie tasks previously sent for compute verification. Once a task $T_i$ is stored in CVT, \sol keeps track of all results received for $T_i$ and the corresponding server nodes returning these results. 

\begin{equation}
    CVT= \{T_1, \ldots, T_L\},  
    \label{eq.cvt}
\end{equation}

\begin{equation}
    T_i=[V^i,R^i_1, R^i_2, \ldots, R^i_N],
    \label{eq.T}
\end{equation}

where $V^i$ is the majority result for task $T_i$ or $NULL$ when there is no majority recorded yet for task $T_i$, and $R^i_j$ is the returned results received by server $S_j$ when performed task $T_i$, or $\O$ when $S_j$ did not receive task $T_i$.

\sol runs the collusion detection module periodically, every period $\Delta t$. It selects a collusion detection task  $T_i \in CVT$ randomly and sends it to collusion detection pool, $C_P=\{S^i_j| R^i_j=\O \}$, consisting of a set of servers that have never received the same task previously (\ie no results returned/recorded by the servers). When \sol fails to find to select $k$ servers that have never received task $T_i$, \ie $|C_P| \leq k$, it removes $T_i$ from CVT and replaces it with the most recent genuine task.

The collusion detection algorithm runs as soon as \sol receives $r^i_j$, a result from server $S^i_j \in C_P$. Collusion is triggered if and only if: (i) the result received does not agree with the recorded majority result, and (ii) there exist another same result returned by a different server (\ie servers agreeing on the same result which is different from the majority result), Alg.~\ref{detectalg} Line~\ref{detectalg-second}.
If collusion is not detected, \ie \func function returns -1, the majority result value is updated if $V^i==NULL$ and the $r^i_j$ is inserted into the CVT, \ie $R^i_j \gets r^i_j$. However, if collusion is detected, \sol wipes the CVT, and immediately initiates the mitigation module to detect and isolate colluding servers.


\begin{figure}[tbp]
    \centering
    \includegraphics[width=1.0\linewidth]{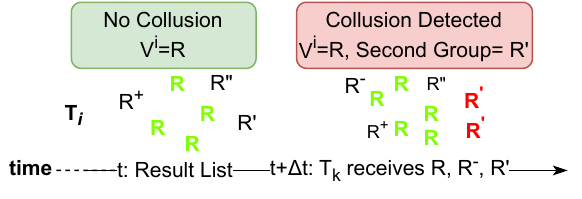}
    \caption{In time t, there is just a group of majority ($V^i=R$), but in $t+\Delta t$, the received result $R'$ makes the second majority group, and collusion detected.}
    \label{CVT}
\end{figure}

Figure~\ref{CVT} depicts an example, where a user sends a task $T_i$ at time $t+\Delta t$, where the received results $R$ and $R''$ did not trigger a collusion detection, however received result $R'$ did, because \sol found another $R'\in CVT$ and $V^i=R\neq R'$, which resulted in finding two separate group of server agreeing on two different set of results, resulting a detection of the presence of colluding servers in the network.    

\begin{algorithm}
\begin{algorithmic}[1]
		\REQUIRE $T_i,r^i_j$     
  \IF{$r^i_j==V^i OR V^i==NULL$}
	\STATE \textbf{return (-1)}\label{detectalg-justignore} \COMMENT{No Collusion}
 \ENDIF
 	\IF{$ResultInCVT(r^i_j)$ AND $V^i \neq r^i_j$}
		\STATE \textbf{return (1)}\label{detectalg-second} \COMMENT{Collusion Detected}
	\ENDIF
 \STATE \textbf{return (-1)} 
\end{algorithmic}
\caption{\sol's \func Function, after receiving $r^i_j$ from server $S_j$}
 \label{detectalg}
\end{algorithm}

\subsection{\sol's Mitigation Module}
As soon as collusion is detected, the mitigation module is triggered to proactively probe a subset of servers in order to accurately classify them into {\em honest} or {\em colluding} workers.


\sol's detection algorithm implies that there is at least two colluding nodes in the network, based on one verification task. The mitigation module consists of (i) clustering the nodes into two groups of servers that have similar behavior, which we refer to as {\em similarity-based grouping}, and (ii) identifying which group includes {\em honest} and which one includes {\em colluding} servers, we refer to this step as {\em identifying colluders}.


\subsubsection{Similarity-based Partitioning into Two Unnamed Groups} \label{grouping-sub}

While detection has found that there may exist two groups of servers, these groups are not exhaustive because they are formed based on a single task and a subset of probed nodes. Therefore, the mitigation module is designed to probe all the nodes in the network and construct an exhaustive undirected weighted similarity graph (SG), based on how often pairs of workers agreed on the same result for the same task (\ie agree on voting outcomes).  

Therefore, \sol collects new votes based on current tasks until each pair of workers have been selected in 8\footnote{It has been shown that 8 is sufficient to have an accurate similarity graph~\cite{Staab09}} separate voting pools. Tasks are sent to only one pool of workers and the task, its $k$ votes, and the $k$ worker nodes returning the vote results are stored into a new task repository, $TR = \{T_i, i=1 \ldots L \mid T_i=\{(R^i_1, S_1), (R^i_2, S_2),(R^i_3, S_3)  \}  \}$ (assuming that the pool size is $k=3$). 

Then SG is constructed as a complete graph with $N$ vertices, and $\frac{N \times (N + 1)}{2}$ weighted edges, where weights are calculated as the ratio between how often two workers', $S_i$'s and $S_j$'s, votes agreed with each others by the number of times they appeared on the same voting pools: \ie the weighted edge connecting two worker $S_i$ and $S_j$, $e_{i,j}$ is: 
$$e_{i,j} = \frac{ \sum_k \bbone_{R^k_i=R^k_j}}{\sum_k \bbone_{\exists(R^k_i \cap R^k_j)}} $$

The Similarity-based Grouping starts by isolating the naive malicious workers which will return results/votes while consistently disagreeing with all other nodes. Isolating these workers is simple; we apply the EigenTrust algorithm proposed by Kamvar
et al.~\cite{kamvar2003eigentrust} to isolate naive malicious workers, which we will save into a list of malicious workers \m. After removing \m from the $SG$ graph, the resulting graph consists only of {\em honest} and {\em colluding} workers.



We use a graph partitioning algorithm to cut the graph into two sub-graphs forming two disjoint groups. Graph partitioning algorithms such as Markov Cluster Algorithm (MCL)~\cite{markovclustering}, Mininmum Cut Tree Clustering (MinCTC)~\cite{brandes2003experiments}, or Spectral Clustering (SP)~\cite{spectral}, can be used. 
We use one of these algorithm attractively and stop as soon as the graph is portioned into two sub-graphs. 

However, if the graph portioning algorithm fails to portions the $SG$ graph, while we have detected the presence of colluding workers in the network (\ie based on the collusion detection module), \sol uses a greedy heuristic to construct two groups of nodes based the outcome of the detection module. Assume the collusion detection module has detected collusion based on task $T_i$ after receiving a voting results $R^i_j$ from worker $S_j$.

Therefore, we construct the two groups as follows; worker node $S_j$ will form a group $G1$ with the other server worker $S_k$ such that $R^i_j=R^i_k$ (\eg from the example of Figure~\ref{detectalg}, G1 will be formed by the two worker nodes returning the red R' result), The other group $G2$ is formed by all worker nodes returning the majority vote result and the remaining worker nodes which did not return any voting result yet (\eg from the example of Figure~\ref{detectalg}, G2 will be formed by the all worker nodes returning the green R result as well as all other worker nodes which where not probed yet). $G2$ will then be updated by removing all naive malicious workers \m computed prior to the group partitioning.




\subsubsection{Group Identification: Identify and Isolate Colluding Workers} \label{Naming-sub}

At this step, the main goal is to identify which group includes honest and which one includes colluding workers. 
{\em Unlike other state-of-the-art research, we do not assume the presence of a trusted third party servers or trusted tasks (\ie with guaranteed results) to guarantee efficient identification of these two groups}.

\sol constructs a subset of trusted tasks ($TT$) from the original task repository, $TR$. A task $T_k \in TR$ is called a trusted task if and only if $\exists i \in G1~ and ~j \in G2 | R^k_i = R^k_j$, where $G1$ and $G2$ are the two unnamed groups identified in the previous step. In other words, trusted tasks are the ones where workers from the two disjoints groups have agreed upon--colluding workers did not collude for these trusted tasks, thus the result returned for this task can be trusted. Note that colluding workers may not decide to collude if they did not form a majority of the pool or with a probability of $1-P_c$. 

The list $TT$ of trusted tasks will therefore be used to classify the workers into honest and colluding workers. However, since colluding nodes may not collude all the time, \sol sends multiple tasks for each worker of one of the two unnamed groups to classify the group, say $G1$, then the remaining group, $G2$, will constitute the other class of workers. We show, in sec.~\ref{eval.mitigation}, that this idea may not be sufficient and we may need to check the two groups instead of relying only on classifying only one and infer the other.





\sol's group identification uses fewer trusted tasks and resources, if it starts the identification of the honest group, rather than the colluding workers group (details will be presented later in this section as we present the algorithm). In other words, if we start identification of $G1$ and $G1$ was classified as an {\em honest group}, verifying $G2$, consisting of colluding nodes is less complex and requires less trusted task. Otherwise, if $G1$ was classified as a {\em colluding group}, then verifying the {\em honest group}, $G2$ requires more trusted tasks.


\sol uses the size of the group to predict the group class, using the assumption that honest workers are most probably more than colluding workers in the network. Note that \sol is also able to correctly classify the worker even when this assumption is not accurate--\ie colluding workers represent the majority of the workers as we will explain in the algorithm and show results in the evaluation section~\ref{eval.mitigation}. 

Say $G1$ is the bigger group, \sol selects a pool of $k$ workers $P \subseteq G1$ such that $S_j \in P$ have the minimum number of verification within $G1$, then for the selected pool $P$, \sol finds the first task $T_i \in TT$ such all workers $S_j \in P$ have never received task $T_i$, \ie find $T_i$ such that $\forall S_j\in P, (S_j, *) \notin TT$. The verification algorithm stops when the minimum number of verification for all nodes is equal to $e$ or when all tasks in $TT$ have been utilized. The parameter $e$ is the maximum number of trusted tasks verification used to tune the algorithm for accuracy.


\sol assigns a reputation score $Rs_i$ for each worker node $S_i$ based on worker node $S_i$'s votes after comparing these votes with the trusted tasks majority vote result. We assign a voting score of 1 if the vote agrees with majority and -1 otherwise. Let $V_i=\{v_1 \ldots v_{e'}\}$ where $e' \leq e$ be the list of voting scores (\ie a list of 1 or -1). $Rs_i$ is computed as follows:  

$$Rs_i= \frac{\sum^{e'}_{j=1} v_j}{e'}$$






If $\forall S_i \in G1 Rs_i=1$ then \sol identify $G1$ as an honest group--all workers in G1 are honest. Otherwise, \sol runs a k-mean clustering algorithm, with $K=2$ to classify $G1$ into two subgroups, $G1.1$ and $G1.2$. Assume that we name $G1.1$ such that the average reputation score of $G1.1$, $\overline{R_{i\in G1.1}}$ is the highest --- $\overline{R_{i\in G1.1}} > \overline{R_{i\in G1.2}}$. If (I) $|G1.1| \ge |G1.2|$, then, $G2 \gets G2 \cup G1.2$, and $G1 \gets G1.1$; $G1$ consists of honest workers and $G2$ consists primarily of colluding workers. Otherwise (II), \ie $|G1.1| < |G1.2|$, then, $G2 \gets G2 \cup G1.1$, and $G1 \gets G1.2$; $G1$ consists of colluding workers and $G2$ consists primarily of honest workers. 

However, \sol needs to verify that all workers $S_j \in G2$ are colluding (in case I or honest in case II) workers. 

\noindent {\em Case (I)}: The unverified unnamed group $G2$ consists primarily of colluding worker nodes. In this case, \sol selects a pool of $k$ workers from the unnamed group $G2$, \ie $P'=\{S_i, i=1 \ldots k \mid S_i \in G2\}$. \sol selects tasks $T_i \in TT$ to verify the new pool $P'$, similar to the selection for pool $P$. Any worker node $S_i$ returning a vote which disagrees with the majority will be placed in the honest group, \ie $G1 \gets G1 \cup {S_i}$, because $S_i$ did not collude with the majority of colluding nodes, thus $S_i$ is honest. We repeat this process until we exhaust all tasks $T_i \in TT$, because colluding nodes may not always collude and must be tested multiple times. 

\noindent {\em Case (II)}: The unverified unnamed group $G2$ consists primarily of honest worker nodes. In this case, \sol iterates over all worker nodes $S_i \in G2$ and selects a pool of $k$ workers formed with $S_i$ and the remaining are colluding nodes, \ie $P'={S_1,S_2,\ldots,S_k}$ such that $\exists i \mid S_i \in G2$ AND $\forall j\neq i, S_j \in G1$. \sol selects up to $e$ tasks $T_i \in TT$ to verify the new pool $P'$, similar to the selection for pool $P$. In this case, if worker node $S_i\in G2 \cap P'$ returns a vote which disagrees with the majority, then $S_i$ is verified as honest and remains in $G2$, \sol then selects/verifies another worker $S_j$ until all nodes are verified.
However, if after $e$ task verification, all worker nodes in $P'$ return the same vote then they are all colluding workers and \sol will place $S_i$ in the colluding group, \ie $G1 \gets G1 \cup {S_i}$.

\begin{figure}[tbp]
    \centering
    \includegraphics[width=1.0\linewidth]{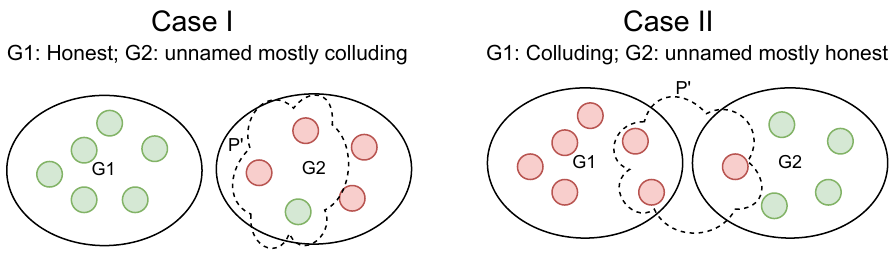}
    \caption{Group identification example: In Case (I), where G1 consists of honest workers, \sol selects pools P' entirely from G2; nodes disagreeing with majority are honest and added to G1; In Case (II), where G1 consists of colluding workers, P' is formed with two nodes from G1 and one from G2, until we verify all G2 members (\ie G2 members agreeing with majority are colluding and added to G1).}
    \label{PC}
\end{figure}

Figure~\ref{PC} depicts two examples of case (I) and (II); in the first example G1 was named as \h thus G2 is unnamed and consists mainly of colluding nodes, however the clustering algorithm may have misclassified few honest workers as clouding. In this case, \sol selects its pool of workers for verification from G2 and look for any inconsistency in the votes which result in classifying the node as honest and adding it to G1. In case (II) (the sub-figure from the right in Figure~\ref{PC}), G2 consists primarily of honest workers, thus \sol verifies worker by worker from G2 with a pool P' formed with colluding workers from G1. A pool showing zero inconsistencies, results in classifying the verified worker as colluding. 


Finally, \sol has identified all worker groups (i) naive malicious (\m), honest workers (\h), and colluding workers (\cl). \sol then frees all task sets, $TR$ and $TT$ and runs the detection module again periodically to find new potentially {\em colluding} nodes.

\section{Evaluation} \label{Evaluation}

In this section, we compare \sol's performance to the closest state-of-the-art algorithm, Staab \& Angel, which we refer to as SnE~\cite{Staab09}, one of the existing collusion mitigation approaches that uses comparable set of assumptions. Note that none of the existing research has claimed or developed a method to detect and mitigate collusion when colluding workers exceeds 50\% of the nodes in the network.

\subsection{Simulation Setup}
We implement \sol and simulate different workload and network scenario using a python simulator we have developed. We consider a network of $N=20$ workers. While we have considered naive malicious workers in our design, we do set \m$=\not0$ because identifying \m is very trivial and has been solved by many state-of-the-art algorithms including \stabb~\cite{Staab09}. We simulate a 20 to 25 milliseconds random round trip time communication delay between all nodes at the edge. We implement two variations of \stabb, \stabb8 and \stabb12 using using $e=8$ (recommended value in~\cite{Staab09}) and $e=12$ observations per edge. We have tested other implementations using different clustering algorithm such as SP and MinCTC, however in this paper we show only results for \stabb using MCL clustering which achieved the best results. 

We vary the percentage of colluding workers \cl from 10\%, to \%90 from the set of $N$ workers. Colluding nodes can collude with a probability  $P_c$ ranging from 10\% (\ie rarely collude), to 90\% (\ie mostly collude). 

Each node generates a set of tasks at a rate of 1000 tasks per second. Each task is sent to $k=3$ workers for task verification. We run the simulation for 100 seconds and we set the collusion to start randomly following a uniform distribution between 3 seconds and 90 seconds. We repeat each simulation 100 times and measure the average or the distribution of \sol's and SnE's results.


\begin{table}[htbp]
  \centering
  \begin{tabular}{|l|c|c|}
    \hline
     \textbf{Parameters} & \textbf{Acronym} & \textbf{Values} \\
    \hline
     Number of worker nodes &$N$  & 20 \\
    \hline
    \% {\em colluding} workers &$|C|$ & $10\% , \ldots, 90\%$ \\ 
    \hline
    Probability of collusion &\pcol & 10, 50, 90 \% \\
    \hline
    Error rate for honest workers &$\epsilon$ & 0.3\% \\
    \hline
    Simulation end time & -- & 100s\\
    \hline
    Maximum observation per edge&  $e$ & 12 obs/edge\\
    \hline
     
     
   Users task generation rate & -- & 1000 task/sec\\
    \hline
   Communication round trip delay  & RTT& $\{20\ldots25\}$ms\\
 \hline
  \end{tabular}
  \caption{Simulation parameter for evaluation of \sol }
  \label{tab:evaluation-parameters}
\end{table}




\subsection{Evaluation Metrics}

In our experiments,  we evaluate \sol's and SnE's performances using the following metrics:

\begin{itemize}
    \item {\em Collusion detection delay:} 
    Measured as the difference between the time when \sol detects the presence of a collusion and the start time of the collusion. The start time of collusion is simulated (\ie known) in our evaluation. We also measure the number of epochs or iterations in addition to the detection delay in seconds.

    \item {\em Collusion detection accuracy:} We use the f1-score ratio of the accurate detection and the falsely detected collusion (\ie either collusion not detected, or falsely detected).

    \item {\em Collusion mitigation accuracy:} 
    We measure the accuracy of \sol's mitigation algorithm as an  f1-score ratio between the number of accurately detected and falsely detected (\ie colluding workers classified as honest or honest classified as colluding workers) colluding workers. 
    
    \item {\em Collusion mitigation latency:} Measured as the difference in time between completing the mitigation (\eg for \sol, this time is the end of the identification of both groups $G1$ and $G2$) and the start of collusion.



    
\end{itemize}

\subsection{\sol's Collusion Detection Performance}

\sol's collusion detection is measured using two main metrics; collusion detection delay and collusion detection accuracy. 

\begin{figure*}
    \centering
    \subfloat[CDF of detection delays in seconds (x-axis in logscale)
    ]
    {\includegraphics[width=0.3\linewidth]{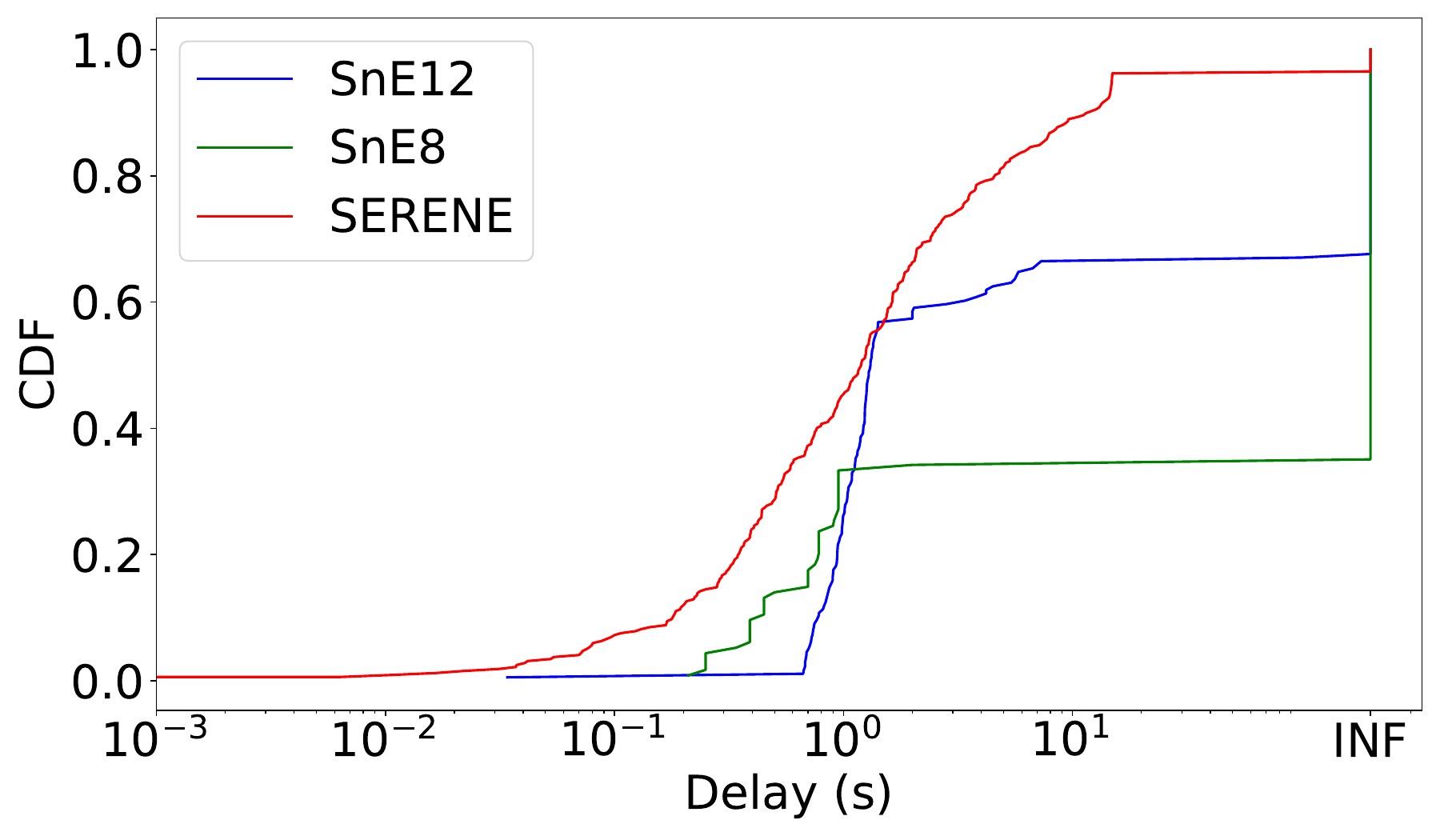}
        \label{evaluation-delaydetectionsecond}}
    \subfloat[CDF of detection delays in epochs (x-axis in logscale)
    ]{\includegraphics[width=0.3\linewidth]{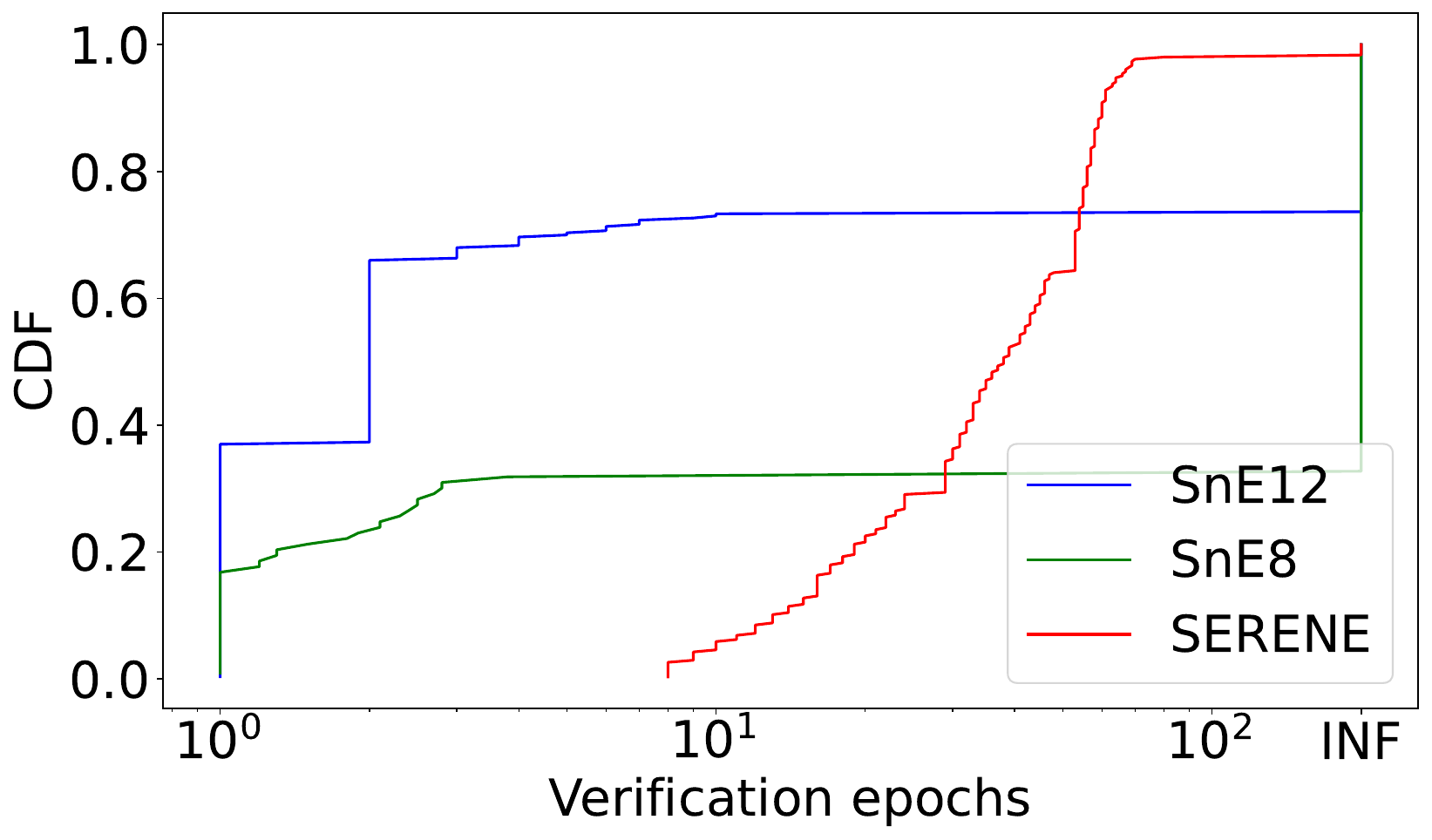}
        \label{evaluation-delaydetectionepoch}}
   \subfloat[Collusion detection accuracy
   ]{\includegraphics[width=0.3\linewidth]{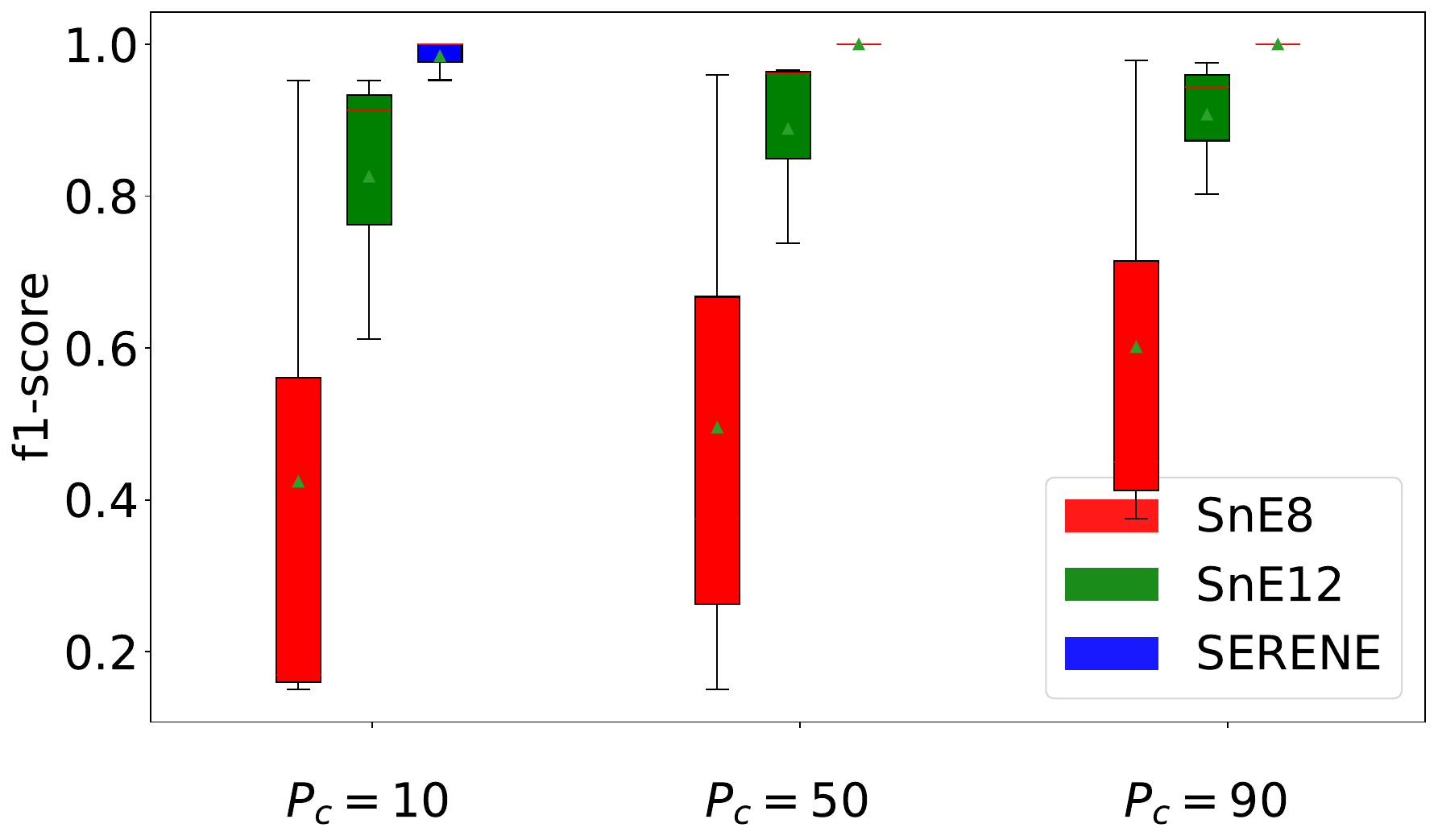}
        \label{evaluation-delaydetectionaccuracy}}
        
    \caption{Comparing \sol's and \stabb's collusion detection delay (a) and (b) and accuracy (c); INF denotes infinite delay values due to unsuccessful collusion detection
    }   
    \label{eval-detection}
\end{figure*}

Figure~\ref{eval-detection} compares the performance of \sol's and \stabb's collusion detection performance with regards to collusion latency (sub-figures a and b) or collusion detection accuracy (sub-figure c). Figure~\ref{eval-detection}-(a) and (b) we measure the detection latency in seconds and in number of iterations respectively and we plot the cumulative distribution function CDF (Inf denotes infinite delays resulting from inability to detect collusion when collusion exists). Note that the CDFs include results with different simulation runs and different \pcol and \cl values.

In Figure~\ref{evaluation-delaydetectionsecond}, \sol outperforms \stabb and detects collusion faster by up to 10 $\times$. For half of the delays (50\% percentile), \sol detects collusion at 0.85 seconds or less however, \sne12 detects collusion at 50\% longer delays, at 1.2 seconds when $e=12$, and mostly unsuccessful delays when $e=8$, \sne8. 
Moreover, while \sol accurately detects collusion with more than 98\%, \stabb fails more than 30\% and 80\% of the time when using $e=12$ and $e=8$ respectively. 

We also plot the CDF of the detection delay in algorithm epochs in Figure~\ref{evaluation-delaydetectionepoch}. Note that the epochs of both \sol and \stabb are incomparable--\ie \stabb operates periodically and waits to gather $e$ observation per edge to perform its clustering, while \sol's epoch is one task at the time and detection occurs when a given task detects two groups. We show that \sol detects collusion accurately in 90 or less tasks with half of the collusion scenario detected within 35 tasks. 

The accuracy of both algorithms is further compared in Figure~\ref{evaluation-delaydetectionaccuracy}. While all algorithms perform better as the probability of colluding increases among the workers, \sol outp erforms \stabb and achieves a 98\% accuracy or more in detecting collusion. However, \stabb, and especially \sne8, show major variation in accuracy performance and fail to detect collusion from 6\% to 27\% of the time for \sne12 and 43\% to 83\% of the time for \sne8, when \pcol=50\%. In fact, \stabb relies on periodic data gathering and triggers a clustering algorithm to find two groups of workers, however most of the time it clusters the network into two group of workers even when there is no collusion in the network. It is worth mentioning that authors did not test \stabb algorithm in absence of collusion in their original paper~\cite{Staab09}.

In addition to \sol's high accuracy performance in detecting colluding workers, the detection algorithm, consisting of zero lookup and simple mathematics operations, is very fast and efficient. We perform a set of bench-marking analysis and show results in section~\ref{eval-benchmarking}.

\noindent{ \bf Impact of CVT size on \sol detection performance:} 
We vary the percentage of collusion $P_c$ to analyze the impact of the size of the CVT table (consists of tasks used to detect the presence of two groups in the network). \sol detects collusion faster as the probability of collusion increases, since the more collusion instances amongst workers the faster we detect two inconsistent groups, thus \sol detects collusion faster. 

\begin{figure}
    \centering
    \subfloat[Collusion detection delay as a function on \pcol ]
    {\includegraphics[width=0.48\linewidth]{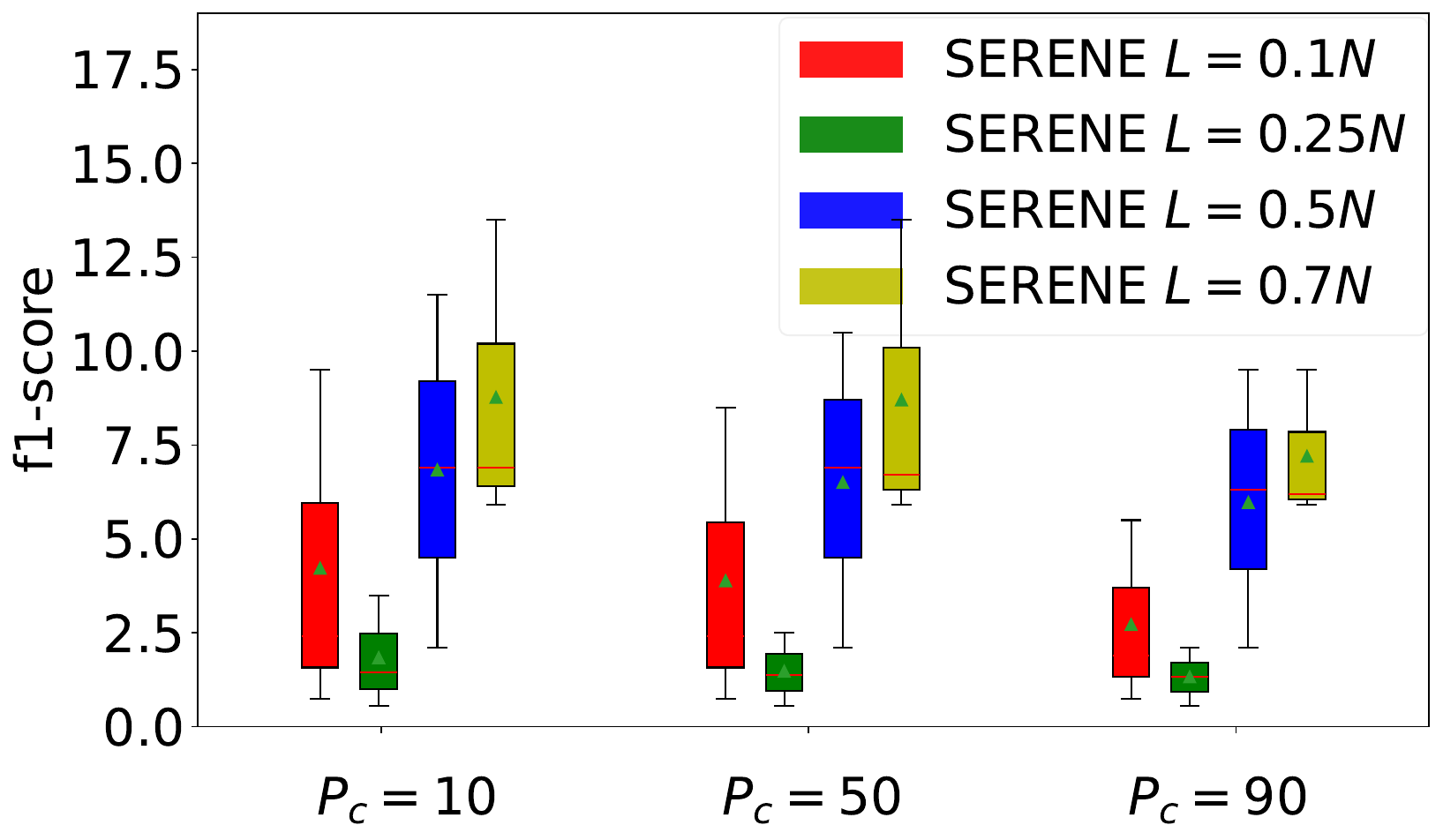}
        \label{evaluation-L1}}
    \subfloat[ CDF of detection delay $\forall$ \pcol, \cl (x-axis in logscale)]{\includegraphics[width=0.48\linewidth]{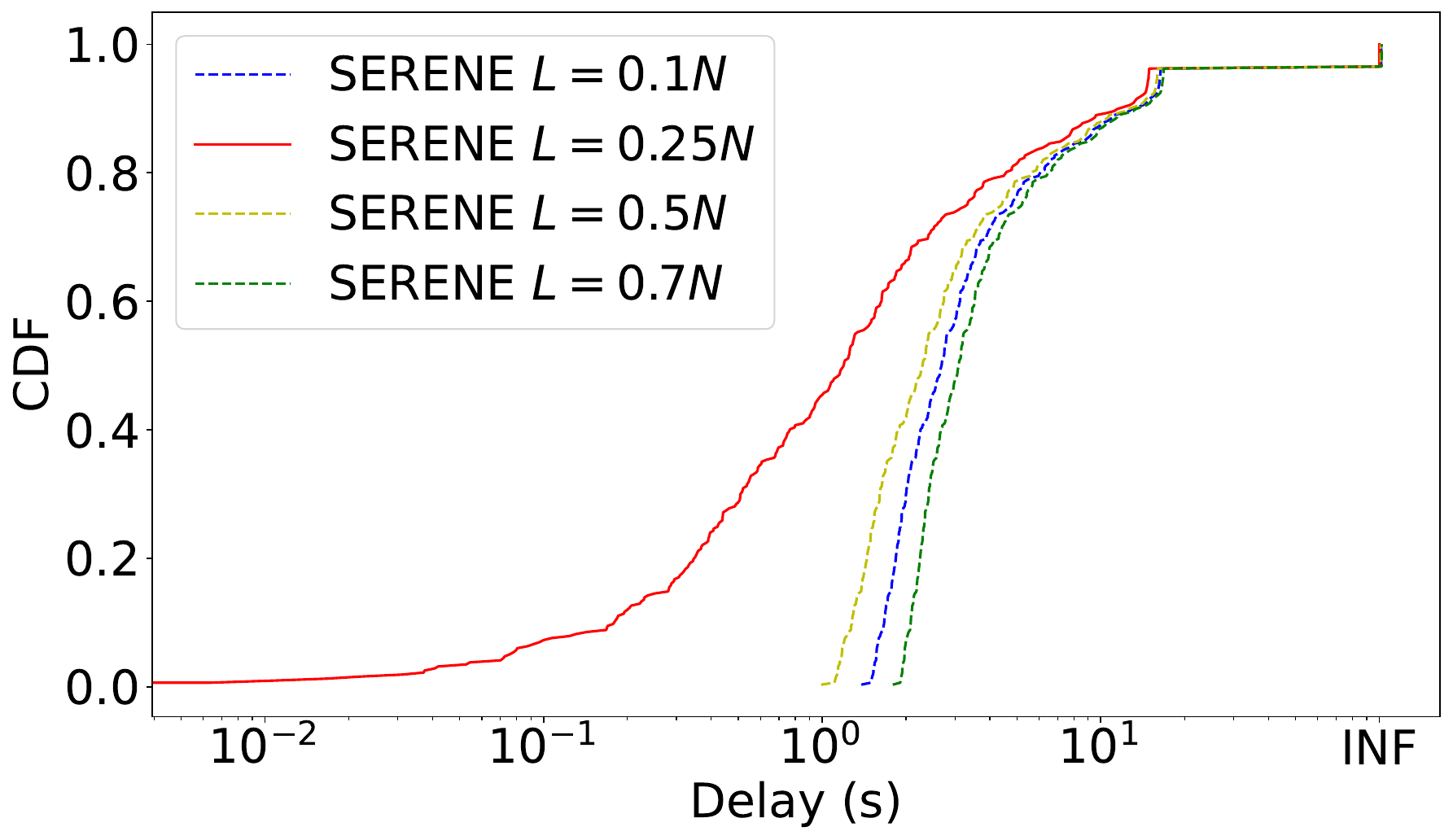}
        \label{evaluation-L2}}
        \caption{Collusion detection delay as a function of  task repository size, $L$}
        \label{fig:L}
\end{figure}

In addition, we show in Figure~\ref{fig:L} that there may exist an optimal value of $L$ ($L\approx .25 \times N$) and very large ($L=.7 \times N$) and very small ($L=.1 \times N$) values perform poorly. Large $L$ values are discouraged because the larger the table size the longer to gather multiple votes for any given task (because \sol chooses tasks randomly), 
thus the slower the detection. On the other hand, very small table sizes results of replacing the tasks that have been used to verify all workers, then \sol replaces them which slows the detection. The CDF of all collusion detection delays $\forall$ \pcol, \cl  (Figure~\ref{evaluation-L2}) shows that $L=.25 \times N$ can achieve more than $10\times$ improvement in latency, making it an important tuning parameter for \sol's collusion detection. 



\subsection{\sol Collusion Mitigation Performance} \label{eval.mitigation}

\begin{figure*}[tbp]
    \centering
    \subfloat[Comparing mitigation accuracy for $P_c=50\%$]
     {\includegraphics[width=0.32\linewidth]{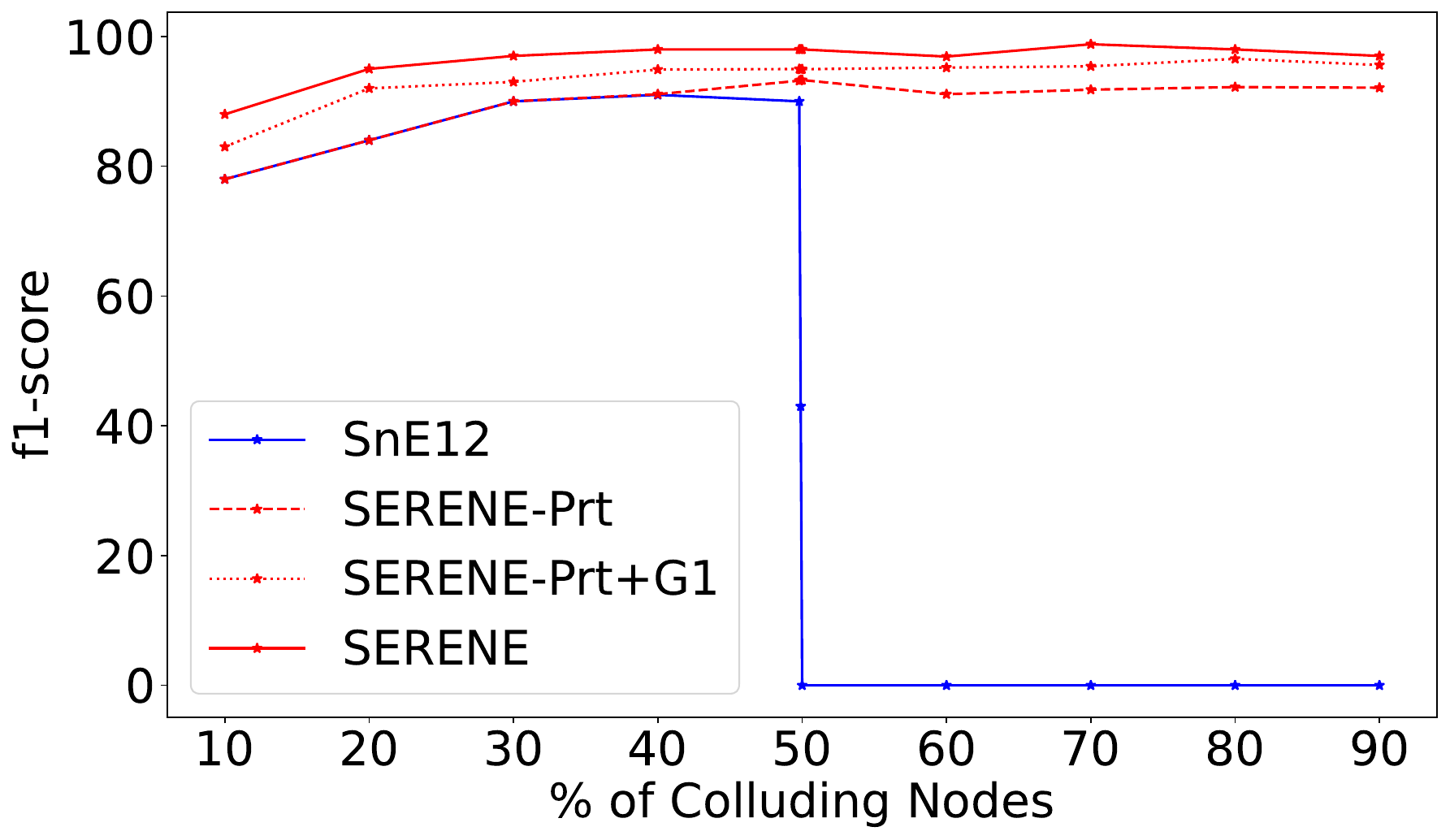}
    \label{f1-population}}
    \subfloat[Collusion mitigation accuracy $\forall$ \pcol]
     {\includegraphics[width=0.32\linewidth]{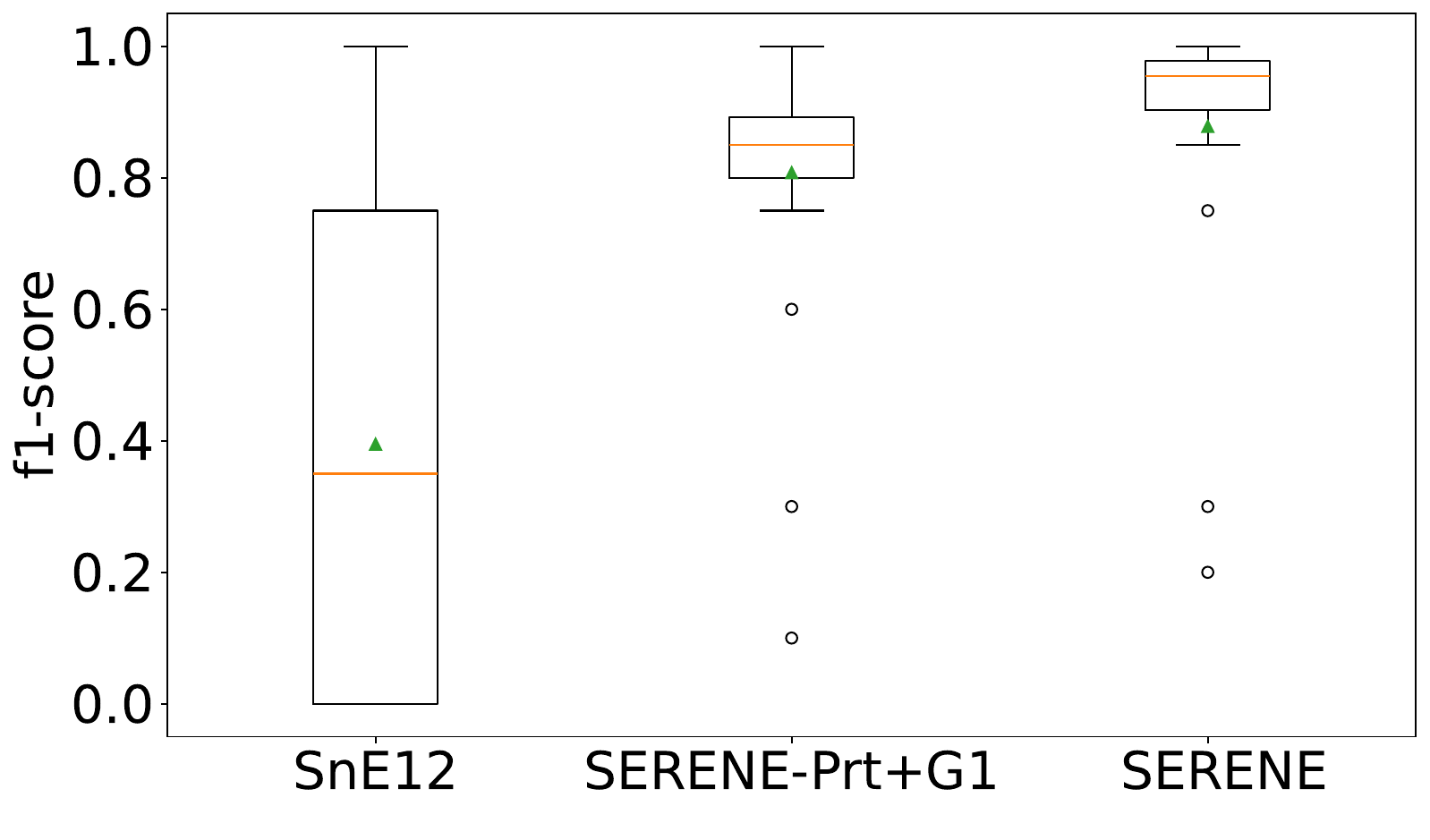}
    \label{f1-all}}
    \subfloat[CDF of mitigation latency for all \cl and \pcol values]
    {\includegraphics[width=0.32\linewidth]{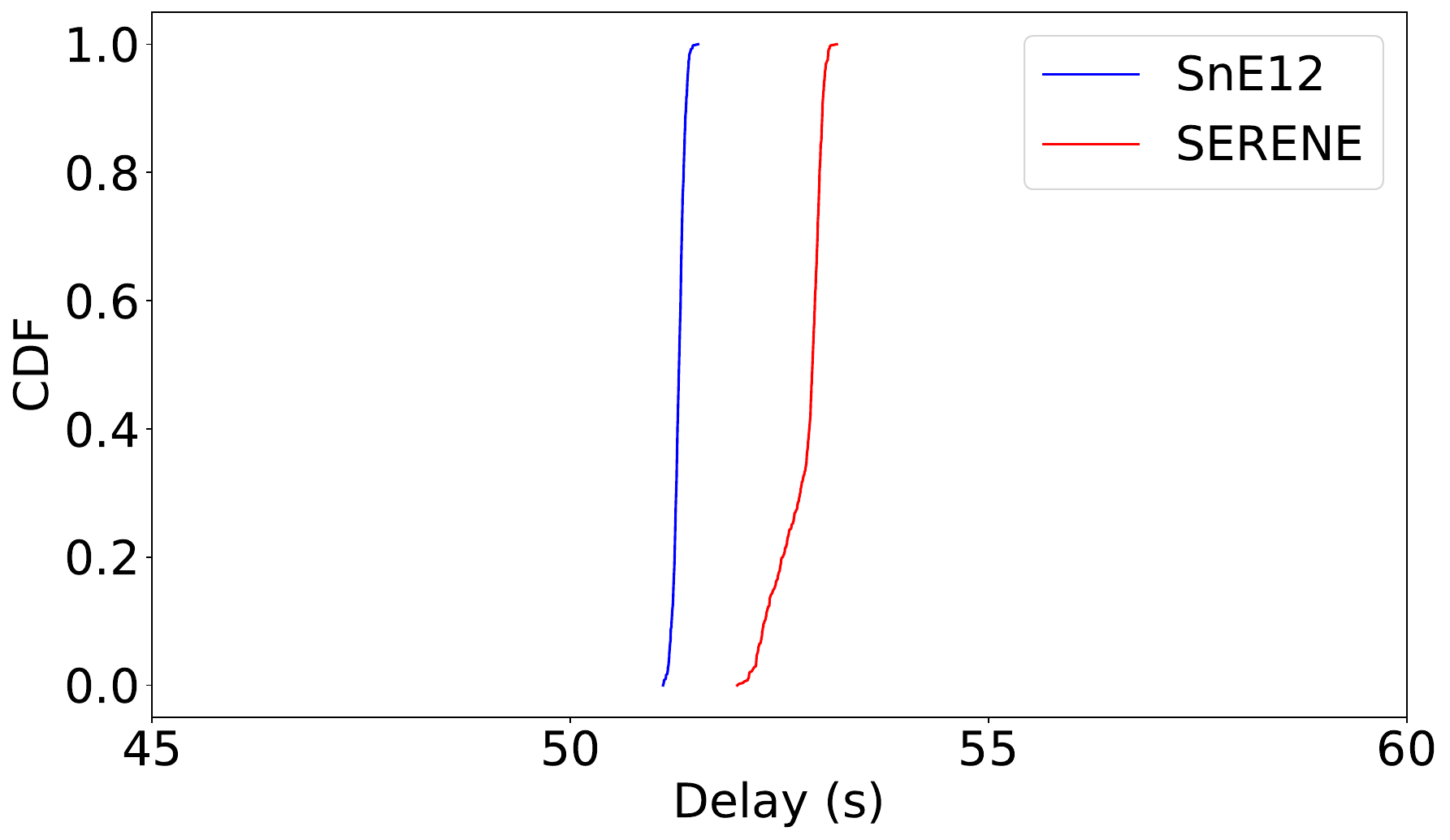}
    \label{mitigation_delay_cdf}}
    \caption{Comparing mitigation accuracy (a, b) and latency (c) for \stabb, \sol implementation up to the grouping phase (\sol-Prt), \sol implementation up to the identification of G1 (\sol-Prt+G1), and \sol (\ie the full design). }
    \label{fig:mitigation}
\end{figure*}

Collusion detection is the first step towards fixing the collusion problem, efficient mitigation is also important. We evaluate, in Figure~\ref{fig:mitigation}, \sol's mitigation performance by measuring both mitigation accuracy and latency and compare these performance to those of \stabb. We compare \sol only to \sne12 which exhibits the best performance compared to \sne8 and for better readability. We also compare multiple implementations of \sol namely: (i) \sol-Partitioning: an implementation of \sol up to the similarity-based partitioning phase where \sol identifies two unnamed clusters G1 and G2 (\ie section~\ref{Naming-sub}), (ii) \sol-Partitioning+G1): an implementation of \sol up to the identification of G1 and removal of misclassified G1 members (\ie section~\ref{Naming-sub}), and (iii) \sol: the full implementation of \sol design. 

We show, in Figure~\ref{f1-population}, that all \sol versions outperform \stabb by up to 10\% when the population of colluding workers is low and more than 95\% more when the percentage of colluding workers exceed 50\%. Note that \stabb works only when colluding workers represents the minority group and it uses this assumption to identify \cl. However, \sol performs best when \cl is large because it detects the two groups and gather more observations per edge faster. 
In addition, while \sol-Partitioning performance is almost identical to \stabb's performance for lower \cl values, the G1 identification phase only helps improve \sol's performance by up to 5\% while maintaining more than 85\% accuracy when colluding workers are the majority in the network. 
\sol achieves up to 95\% accuracy when \pcol=0.5, however Figure~\ref{f1-all} shows similar performance when the distribution of delays for all \pcol values in included--\sol outperforms \sne12 and achieves consistent accuracy above 80\% $\forall$ \pcol and \cl. 

This major accuracy gain comes with a minor latency cost shown in Figure~\ref{mitigation_delay_cdf}. While \stabb have almost constant mitigation delay which constitute the time needed to gather enough observation per edge to cluster the graph, \sol uses the identification phase to identify both groups G1 and G2 (\sol does not assume that colluding workers represent the majority in the network) and further verify the correctness of the clustering (\eg misclassified workers). These steps help achieve the accuracy gains highlighted in Figure~\ref{f1-population}. \sol's mitigation latency are 500ms to 1.5s more than \sne12 mitigation delays. We argue that accurate identification of colluding workers in the network is essential, thus the additional delay cost introduced by \sol is justified; in fact, \stabb with inaccurate identification of colluding nodes may use another detection and mitigation phased to identify the misclassified workers which will result in much larger latencies.


\begin{figure}[tbp]
    \centering  
   \includegraphics[width=0.8\linewidth]{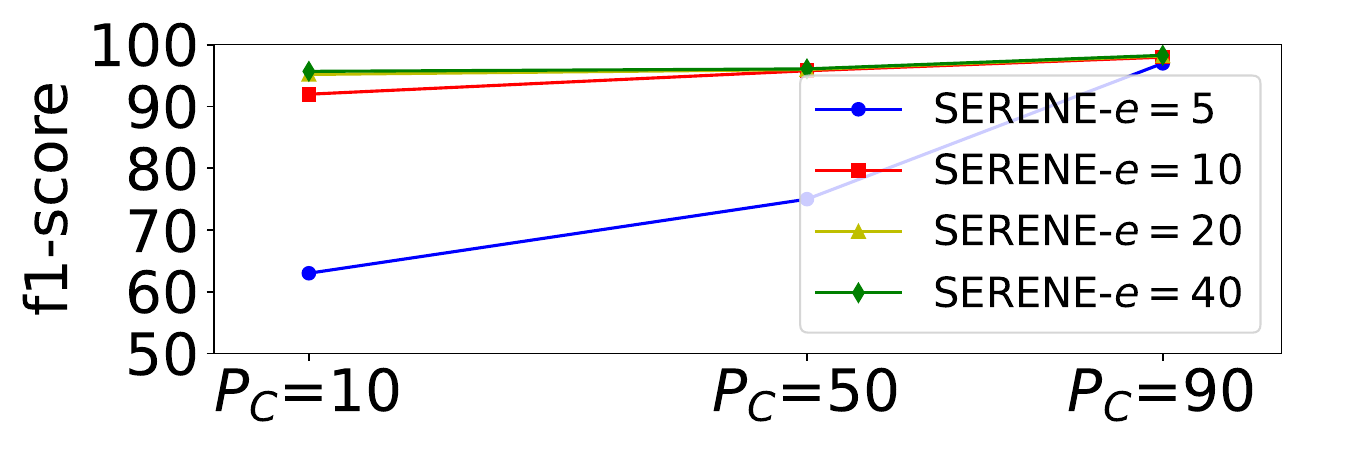}
    \caption{Comparing the collusion mitigation accuracy for different observation per edge values ($e$).}
    \label{mitigation-e}
\end{figure}

\noindent {\bf Effect of $e$ on collusion mitigation accuracy:}
We have discussed the impact of $e$ on \stabb performance. \sol can also be tuned with different values of $e$ to increase its mitigation accuracy (detection accuracy does not seem to be impacted by $e$). We compare, in Figure~\ref{mitigation-e}, the collusion mitigation accuracy of \sol when $e=5,10,20,40$. We show that the impact of $e$ is highlighted more for lower probability of collusion, \pcol (more 30\% accuracy gain when comparing $e=5$ and $e=10$). Lower \pcol requires more observations for colluding nodes to consistently collude with each others. However, we also observe that \sol-e=10 seems to achieve a good trade-off and there is not much improvement recorded as we increase $e$ (less than 1.5\% for higher \pcol values). Note that we choose $e=12$ for all other experiments which achieves good trade off between accuracy and delay.    



\subsection{Prototyping and Benchmarking Tests}
\label{eval-benchmarking}
We perform a set of benchmarking tests using different machines/platforms including: (i) a Raspberry Pi v3b with Quad Core 1.2GHz 64bit CPU, and 1GB RAM (Pi3), (ii) Raspberry Pi v4 with Quad-core 1.8Ghz CPU, and 4GB of RAM (Pi4), and (iii) an old laptop with Intel dual core 2.0Ghz 64bit CPU, and 2GB RAM (Laptop).

\begin{figure}[tbp]
    \centering  
   \subfloat[Memory and CPU utilization]
     {\includegraphics[width=0.48\linewidth]{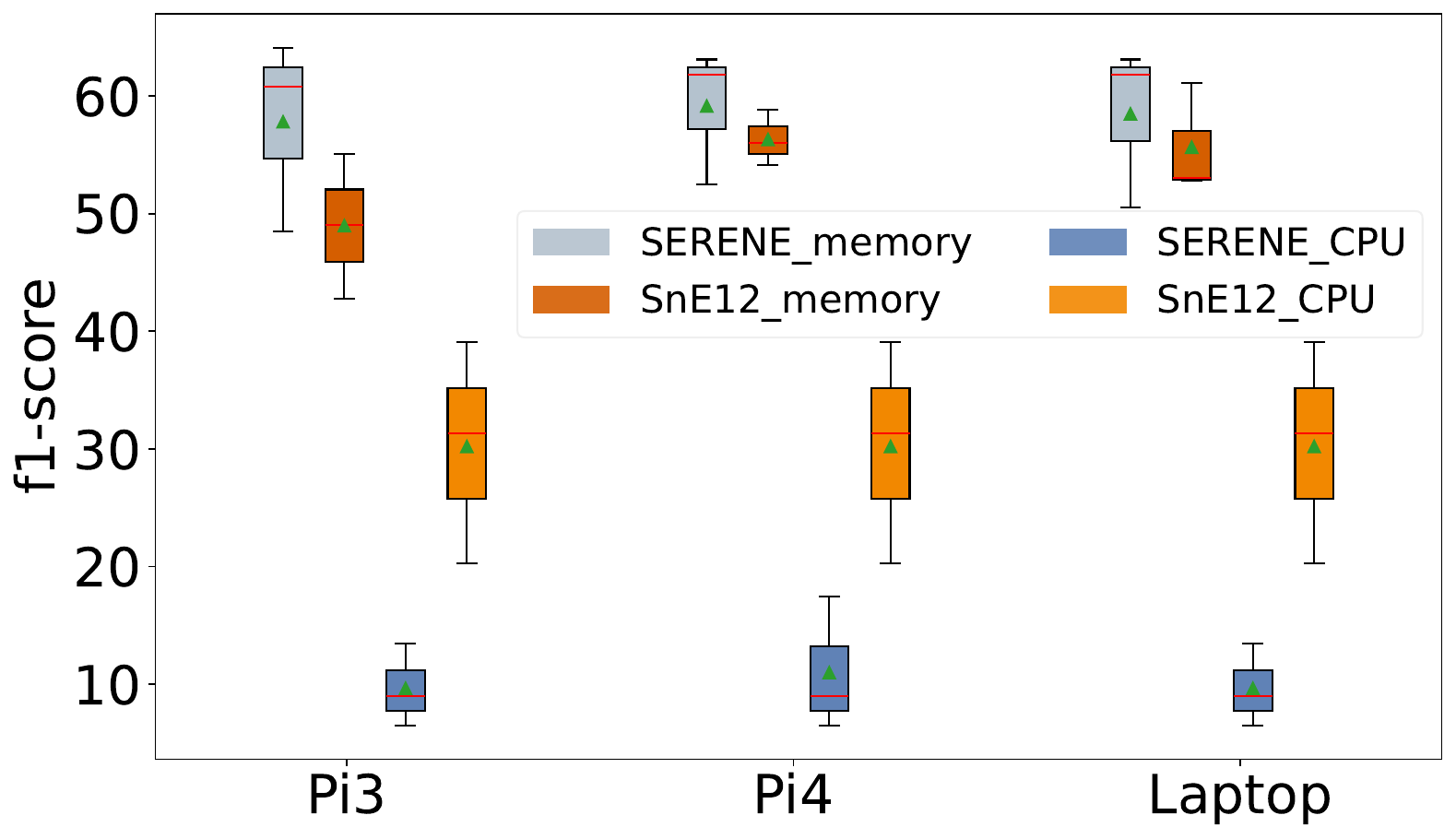}
    \label{benchmarking-cpu}}
     \subfloat[Run times]
     {\includegraphics[width=0.48\linewidth]{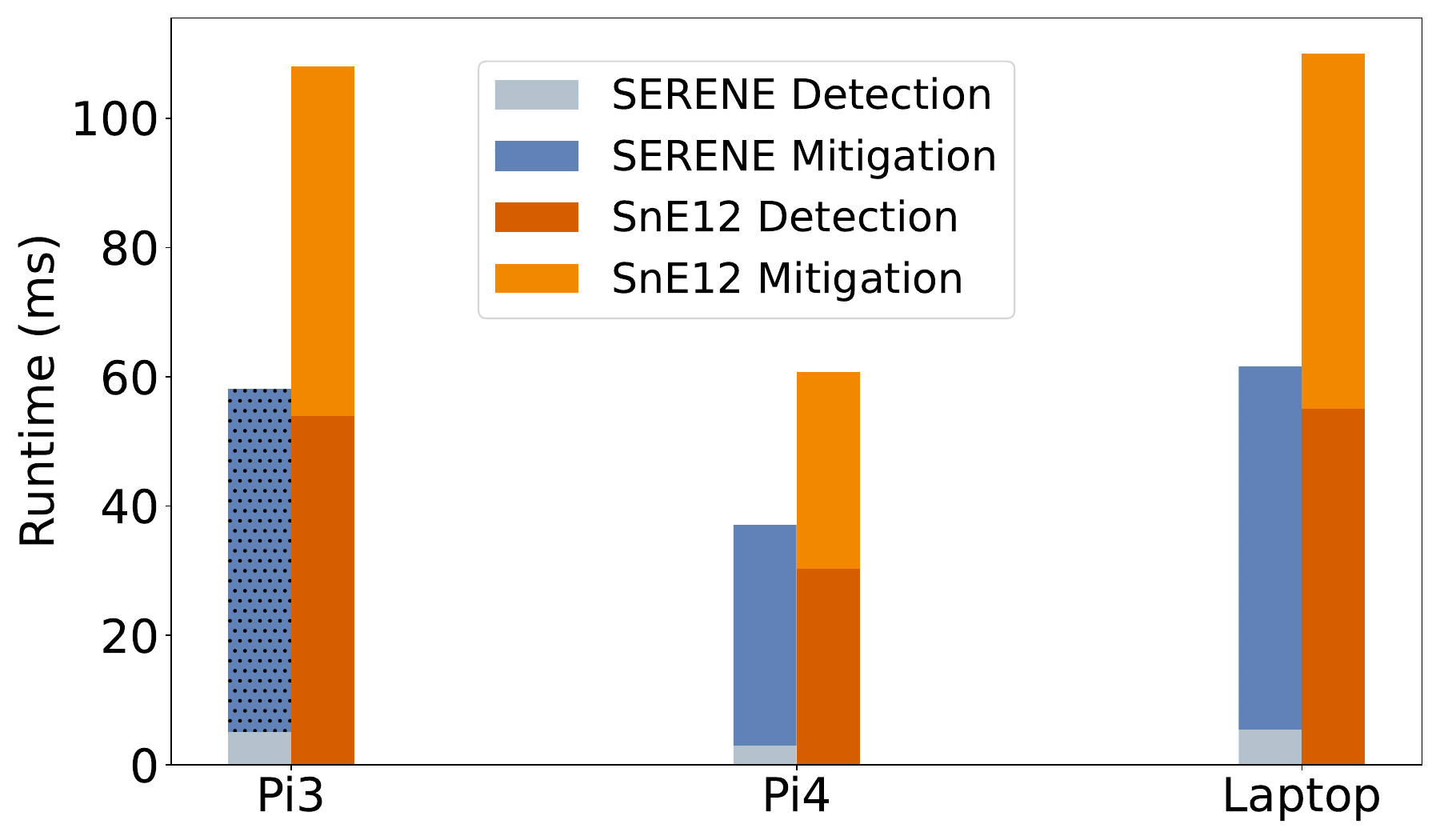} \label{benchmarking_Run}}
    \caption{The benchmarking tests for memory and cpu utilization (a) as well as run times (b) using a Raspberry Pi3, Raspberry Pi4, and a Laptop machine. 
    }
   \label{benchmarking}
\end{figure}

We implement both \sol and \stabb on these three platforms and perform a set of benchmarking tests to measure the CPU usage, memory usage and run times of the proposed schemes. We repeat the experiments 50 times and plots the distribution of all gathered result data in Figure~\ref{benchmarking}. 

While \sol consumes 9\% more memory to load its tables and runs the detection and mitigation algorithms (Figure~\ref{benchmarking-cpu}), it uses more than half the CPU when compared to \stabb's CPU usage--\sol uses on average 10\% CPU while \stabb uses 33\%. Note that none of the tested platform shows over-utilization of the resources, thus the very small differences when we compare results across the three platforms. 

However, we can show in Figure~\ref{benchmarking_Run}, that \sol runs two times faster than \stabb across the three tested platforms. This gain is mainly due to the fast detection algorithm which runs $10\times$ faster than \stabb's collusion detection-- \sol detects in 5 milliseconds, while \stabb detects collusion in 54 milliseconds. Note that the two algorithm were implemented in the same fashion--\ie no parallel computing for \sol or \stabb.




\section{Discussion, Limitations, and Future Directions} \label{limitation}
In this section, we discuss about some limitations of \sol, and challenges that can be addressed. 

\begin{itemize}
    \item We have assumed a fixed graph of nodes representing a list of workers that do not change over time. With mobility, for instance vehicles, must construct dynamic graphs and remove nodes that has not been probed in a period of time. This temporal graph can be more complex to analyze, store, and manage. Note than none of the state-of-the-art solutions have studied this challenging problem. We plan to extend \sol with a smart dynamic temporal graph partitioning algorithm to detect and mitigate the collusion in a mobile environment.

    \item We have studied the threat model where all colluding nodes at the edge coordinate and return the same incorrect result. In reality, there maybe different clusters of colluding workers operating simultaneously. While this helps detect collusion faster as \sol will be able to detect multiple groups of workers, the mitigation process needs tuning to accommodate this scenario.  

    \item While we argue that replication-based task verification are more generic and can be applied to any edge computing application, comparing \sol to non-replication based solutions such in TEE or crypto-based solution can be further investigated. We can run a set of benchmarking tests using real-world workloads and investigate the accuracy, delay, and overhead across all these categories of solutions.  
    

\end{itemize}

\section{Concluding Remarks} 
\label{colclusion}
We have presented a new worker collusion resilient replication based task verification scheme called \sol. Unlike state-of-the-art collusion resilient solutions, \sol is lightweight and is able to efficiently and quickly detect the presence of colluding workers in the network, then isolate them.  

\sol's detection relies on identifying two cluster of workers consistently disagreeing with each others. While this identification guarantees the presence of colluding workers however, without assumptions of the size of colluding workers or the presence of trusted third party servers (used by state-of-the-art solutions), \sol uses a three step mitigation to partition the group of workers and identify the colluding ones.  
Our results show that \sol detects the existence of collusion more accurately (with more than 98\% success ratio, which represent 30\% improvement compared to \stabb), and $5\times$ faster than \stabb's detection delay. Moreover the \sol's mitigation accuracy is 10\% better that state-of-the-art and when the number of colluding workers exceeds 50\% of the total worker population \sol's maintains a 90-95\% accuracy why \stabb fails and achieves 0\% success in identifying colluding workers. 




In the future, we plan to improve \sol's implementation to optimize its resource utilization (storage, and lookup time), as well as the number of messages sent (overhead) in the mitigation process. In addition, we would like to test a real-world prototype and deployment of \sol in an edge computing setting with real-world workloads.  


\section*{ACKNOWLEDGMENTS}
This work was partially supported by US NSF award \#2148358, and an UMSL Research Award.



   






\bibliographystyle{ieeetr}
\bibliography{main}
\end{document}